 \newlength\smallfigwidth
\documentclass[aps,prb, 10pt,twocolumn,floatfix,showpacs,amsmath,amssymb,superscriptaddress] {revtex4-1}

\usepackage{color}
\usepackage{amsmath}
\usepackage{graphicx}
\usepackage{graphics}
\usepackage[pdf]{pstricks}
\usepackage{pdftexcmds}
\usepackage{ifpdf}
\usepackage{amssymb}
\usepackage{esint}

\makeatletter
\@ifundefined{textcolor}{}
{%
 \definecolor{BLACK}{gray}{0}
 \definecolor{WHITE}{gray}{1}
 \definecolor{RED}{rgb}{1,0,0}
 \definecolor{GREEN}{rgb}{0,1,0}
 \definecolor{BLUE}{rgb}{0,0,1}
 \definecolor{CYAN}{cmyk}{1,0,0,0}
 \definecolor{MAGENTA}{cmyk}{0,1,0,0}
 \definecolor{YELLOW}{cmyk}{0,0,1,0}
 }

\allowdisplaybreaks
\makeatother

\def \nn{\nonumber}
\newcommand{\Sp}{\mathbf{S}}
\def\vr {{\bf r}}
\def\vk {{\bf k}}

\def\vq {{\bf q}}
\def\vkc {{\bf k}^{*}}

\def \ba {\begin{eqnarray}}
\def \ea {\end{eqnarray}}

\begin{document}

\title{Three-sublattice Skyrmion crystal in the antiferromagnetic triangular lattice}

\author{H.D. Rosales}
\affiliation{Instituto de F\'isica de La Plata and Departamento de F\'isica,
Universidad Nacional de La Plata, C.C. 67, 1900 La Plata, Argentina}
\author{D.C. Cabra}
\affiliation{Instituto de F\'isica de La Plata and Departamento de F\'isica,
Universidad Nacional de La Plata, C.C. 67, 1900 La Plata, Argentina}
\affiliation{Abdus Salam International Centre for Theoretical Physics, Associate Scheme,
Strada Costiera 11, 34151, Trieste, Italy}
\author{Pierre Pujol}
\affiliation{Laboratoire de Physique Th\'eorique, IRSAMC, CNRS and Universit\'e de Toulouse, UPS, F-31062 Toulouse, France}

\date{\today}

\begin{abstract}
The frustrated classical antiferromagnetic Heisenberg model with Dzyaloshinskii-Moriya (DM) interactions on the
triangular lattice is studied under a magnetic field by means of semiclassical calculations and large-scale Monte Carlo simulations.
We show that even a small DM interaction induces the formation of an Antiferromagnetic Skyrmion crystal (AF-SkX) state. Unlike what
is observed in ferromagnetic materials, we show that the AF-SkX state consists of three interpenetrating Skyrmion crystals
(one by sublattice), and most importantly, the AF-SkX state seems to survive in the limit of zero temperature. To characterize the phase
diagram we compute the average of the topological order parameter which can be associated to the number of topological charges or Skyrmions.
As the magnetic field increases this parameter presents a clear jump, indicating a discontinuous transition from a
spiral phase into the AF-SkX phase, where multiple Bragg peaks coexist in the spin structure factor. For higher fields, a second (probably continuous) transition occurs into a featureless paramagnetic phase.
\end{abstract}

\maketitle

\section{Introduction}
Twisted modulated magnetic textures due to an antisymmetric spin exchange interaction,  termed the Dzyaloshinskii-Moriya interaction,
have attracted much interest mainly after the experimental observation of non trivial magnetic configurations, called magnetic Skyrmion
lattices, which have important potential technological applications\cite{ScienceSkx1}.  Recently, a strong evidence of the formation of
a Skyrmion crystal (SkX) state was observed in metallic ferromagnet MnSi\cite{Mulbauer2009}. This state consists of a triangular lattice
arrangement of Skyrmions and can be visualized as a superposition of three non-equivalent spirals states (each characterized by one wavevector
$\vk$) \cite{Bogdanov,Nature2006,Nagaosa2009,Solenov2012,Kanazawa2011,Nature2011}.
In systems with a square lattice structure, the SkX phase arises from the
competition of the ferromagnetic and the Dzyaloshinskii-Moriya (DM) interactions and it is stabilized by a magnetic field and thermal fluctuations.
Recently, Okubo {\it et.al} have shown that it is also possible to stabilize the SkX phase in the isotropic Heisenberg model in the triangular
lattice with strong nearest-neighbor (ferromagnetic) and weak next nearest-neighbor (antiferromagnetic) interactions\cite{Okubo2012}.

In this paper, based in analytical approximations and Monte Carlo simulations,  we show that in a pure antiferromagnetic frustrated system it is
possible to stabilize a phase closely related to the SkX phase described above. The novelty resides in the fact that the state found in the present paper consists of
three interpenetrating Skyrmion lattices (one by sublattice) and, most importantly, this antiferromagnetic SkX (AF-SkX) state survives in the limit $T\to 0$.
Unlike what happens in the ferromagnetic case where it is necessary to include large values of $D/|J|$, in the antiferromagnetic case small values of $D/J$  can stabilize the AF-SkX state.
To identify and characterize this phase we compute a chiral order parameter which determines the density of Skyrmions present in the lattice. Due to the discrete nature of the model,
in the region of large magnetic fields a series of discrete jumps in the order parameter is observed.

The paper is organized as follows: In Sec. \ref{sec-model} we present the Hamiltonian and study the classical ground state solutions at zero magnetic field and zero temperature based on the spherical approximation.
This analysis shows that the ground state has a three-fold degeneracy which could be a source of exotic ordered states that might be realized under applied fields, e.g. various types of states where multiple
wave vectors coexist (so-called multiple-q states). Sec. \ref{sec-mc-sim} contains our Monte Carlo results. We found a quite rich low temperature behavior of the system as the magnetic field is varied. The system goes from a spiral low field phase to an antiferromagnetic Skyrmion lattice phase at higher field.
Finally, increasing the magnetic field further, the system enters into a featureless paramagnetic phase. We pay particular attention to the undoubtedly most interesting phase which is the antiferromagnetic Skyrmion lattice phase. It has the particularity of being composed of three interpenetrating and shifted Skyrmion lattices that realize on each sublattice of the triangular lattice. The MC study is complemented in Sec. \ref{sec-phenomelological} by a phenomenological analysis and in Sec. \ref{sec-OutEquil} we analyze the stability of the Skyrmion phase and briefly discuss the relation of the present study with the square lattice analog. We conclude in
Sec. \ref{sec-Conclusions}  with a summary and discussion of our results.

\section{Model and Spherical Approximation}
\label{sec-model}

Let us consider the antiferromagnetic Heisenberg model on the triangular lattice
in a magnetic field, with the Hamiltonian given by

\begin{eqnarray}
\mathcal{H}&=&J\sum_{\left\langle \vr,\vr'\right\rangle}\Sp_\vr\cdot\Sp_{\vr'}+D\,\delta\hat{\vr}\cdot(\Sp_\vr\times\Sp_{\vr'})-h\sum_\vr S^{z}_{\vr}\quad
\label{eq:Hamiltonian}
\end{eqnarray}
where the spin variables are unimodular classical vectors, $|\Sp_{\vr}|=1$, $J>0$ and $D$ are respectively the antiferromagnetic and DM
couplings, $\delta \hat{\vr}=(\vr'-\vr)/|\vr'-\vr|$ is a unitary vector pointing along the axis, $\sum_{\langle \vr,\vr' \rangle}$  means the sum over
nearest-neighbors (NN) couplings on the triangular lattice with primitive translation vectors of the direct lattice $\vec{e}_1=(1,0)$ and
$\vec{e}_2=(1/2,\sqrt{3}/2)$ and $h$ the strength of the magnetic field along the $z$ axis. The DM interaction is chosen in the direction of
the $\delta\hat{\vr}$ vectors, so as to give rise to spiral spin states with the spins lying in a plane perpendicular to the propagation vector\cite{Nagaosa2009}.

The magnetic phase diagram for the model defined by Eq.(\ref{eq:Hamiltonian}) with $D=0$ has been discussed in \cite{PapersTriangular1,PapersTriangular2}.  At $T=0$  and for zero magnetic field, the system orders in a $120^\circ$ three-sublattice magnetic structure
described by the wave vector $\vk=(4\pi/3,0)$. In a magnetic field the classical energy is minimized for spin configurations constrained
by the magnetization of each triangular plaquette:
\begin{equation}
\Sp_{\bigtriangleup} = \mathbf{h}/(3J) \ .
\label{Triangle}
\end{equation}
This constraint leaves undetermined the orientation of the spin plane and sublattice directions inside that plane. This degeneracy persists
up to the saturation field $h_{s}=9J$. However, due the order-from-disorder effect\cite{VillainOBD} this degeneracy is lifted by thermal fluctuations, which select a kind of states (collinear and coplanar)
over the non-coplanar ones at low temperature. The resulting phases depend on the magnetic field strength: for low field a coplanar so-called Y-state, with one spin pinned in the negative $z$ direction and two canted up, at exactly $h/J=3$ the pseudo-plateau state (at $M=1/3$) with the collinear configuration up-up-down and in the high field region a coplanar canted version of it, which smoothly interpolates with the fully polarized state at $h/J=9$.

\begin{figure}[htb]
\includegraphics[width=8cm]{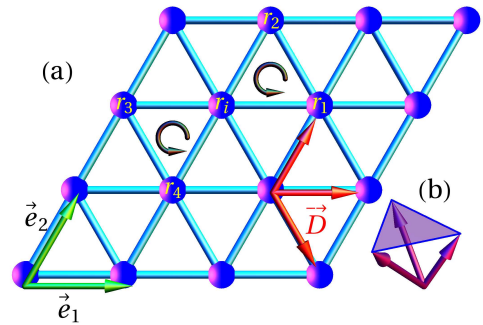}
\caption{(Color online) (a): Triangular lattice. The green arrows indicates the primitive translation vectors of the direct lattice
 $\vec{e}_1=(1,0)$, $\vec{e}_2=(1/2,\sqrt{3}/2)$. Red arrows the Dzyaloshinskii-Moriya vectors. The sites with labels $\vr_i,\vr_1...\vr_4$
 indicate the sites involved in the calculation of the local chirality. (b) Area of the triangle to compute for the discretized Skyrmion number $A^{ab}_{\vr}$ (Eq. (\ref{eq:ChiraOrderPara0})).}
\label{fig:lattice}
\end{figure}

Now we turn-on the DM interaction with the orientation along the nearest neighbors bonds (Fig. \ref{fig:lattice}). As a first step we investigate the magnetic order of the Hamiltonian Eq. (\ref{eq:Hamiltonian}) at zero magnetic field and zero temperature by means of the spherical approximation\cite{NagayimaBook}. Within this
scheme, instead  of imposing the local length constraint $|\Sp_\vr|=1$, one imposes  a milder condition,  $\sum_{\vr}|\Sp_\vr|^2=N\,S^2$,
where $N$ is the number of lattice sites. With this softer constraint, the model Hamiltonian (\ref{eq:Hamiltonian}) can be diagonalized by
a simple Fourier transformation  $S^{\alpha}_{\vr,a}=\sum_{\vk}S^{\alpha}_{\vk,a}e^{i\,\vr\cdot\vk}$. Here, the index  $\alpha=x,y,z$ is the spin component, $a=1,2,3$
is the sublattice label and $\vr$ and $\vk$ denote the position and pseudo-momentum respectively. The Hamiltonian becomes

\begin{eqnarray}
\mathcal{H}&=& \sum_{\vk}\Psi_{-\vk}\cdot \text{M}(\vk)\cdot\Psi_{\vk}\\
\Psi_{\vk}&=&\{S^{x}_{\vk,1},S^{y}_{\vk,1},S^{z}_{\vk,1},S^{x}_{\vk,2},S^{y}_{\vk,2},S^{z}_{\vk,2},S^{x}_{\vk,3},S^{y}_{\vk,3},S^{z}_{\vk,3}\}\nn
\label{eq:Hamiltonian-SA}
\end{eqnarray}
where the $9\times 9$ matrix $\text{M}(\vk)$
\begin{equation}
\text{M}(\vk)=\left[ \begin{array}{ccc}
0 & m_{12} & m_{13} \\
m^{*}_{12} & 0 & m_{32}^{*} \\
m^{*}_{13} & m_{32} & 0 \end{array} \right]
\label{eq:matrixK}
\end{equation}
and the $3\times 3$ matrices $m_{ab}$ depend on $J,D$ and $\vk$ with explicit expressions given in Appendix \ref{AppendixA}.

\begin{figure}[htb]
\includegraphics[width=8.6cm]{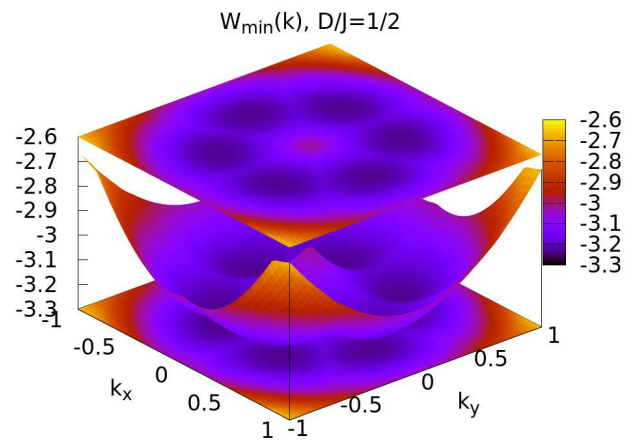}
\caption{(Color online) Minimum eigenvalue $w^{\text{min}}(\vk)$  in the spherical approximation for $D/J=1/2$. For any value of $D/J>0$ we
find that the ground state corresponds to a triple-$q$ state by sublattice.}
\label{fig:wk-sphericalApp}
\end{figure}

 Since we are at $T=0$, the ground-state is obtained from the minimum eigenvalue $w^{\text{min}}(\vk)$ of the matrix M$(\vk)$ (Eq. (\ref{eq:matrixK})).
In particular, we find that for any value of $D/J>0$ there are three minima  (see Fig. \ref{fig:wk-sphericalApp}) suggesting the presence of multiple-q states.
In this configuration the ordering wavevectors $\vkc$ appear  along  the directions of the nearest-neighbor bonds with module
\ba
|\vkc|\simeq\frac{D}{J}+\frac{\sqrt{3}D^2}{4\,J^2}-\frac{D^3}{4\,J^3}-\frac{7\sqrt{3}D^4}{32J^4}+O(\left(\frac{D}{J}\right)^5)
\label{eq:kc}
\ea

In particular we are interested in the non trivial triple-q state, which consists of a superposition of three spirals (each characterized by
one wavevector $\vkc$), that corresponds to the Skyrmion lattice phase. Within the spherical approximation, performed at both zero temperature and zero magnetic field, we find it conceivable that, at very low temperatures, one may neglect all wave vectors other than the three critical modes
$\vkc$ and the uniform $\vk=0$ mode. Henceforth, we expect that it maybe possible to find ordered states characterized by the number
of wave vectors.  In order to elucidate if multiple-q states could be stabilized at finite temperature and magnetic field we explore the behavior of the system by means Monte Carlo simulations in the next Section.

\section{Monte-Carlo simulations}
\label{sec-mc-sim}
\begin{figure*}
\centering{}
\includegraphics[width=2.05\columnwidth]{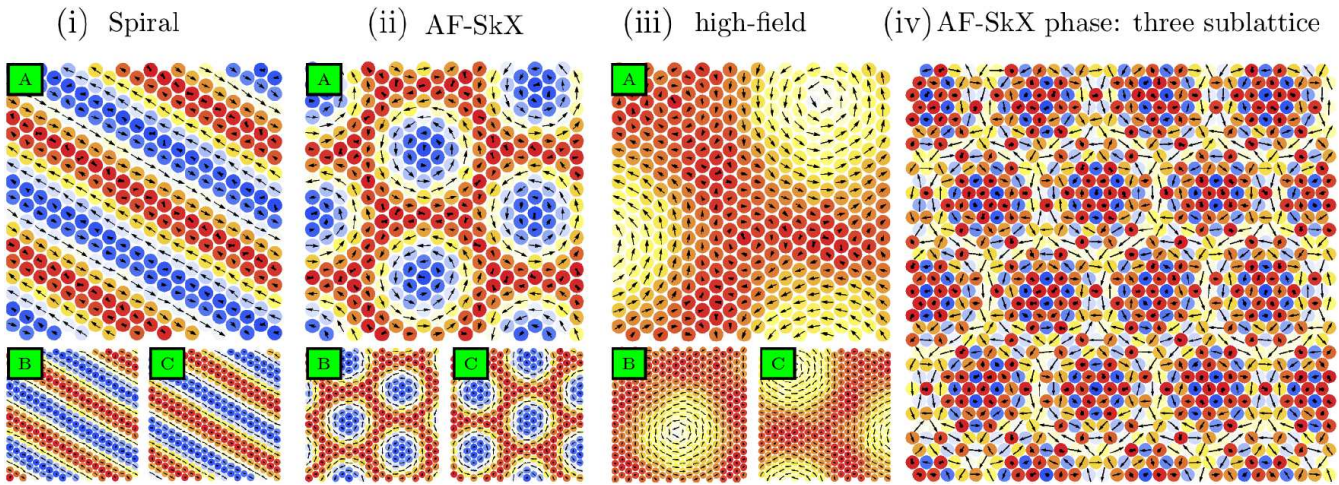}
\caption{(Color online) Snapshots by sublattice for $L=48$, $D/J=1/2$, $T/J=9\times 10^{-3}$ and $h/J=1 (i), 2.4 (ii)$ and $6.2 (iii)$. The green box indicates which
sublattice is ploted (A, B or C). In figure $(iv)$ we show all three sublattice together.	
\label{fig:SnapsN48}}
\end{figure*}

Monte Carlo simulations are performed based on the standard heatbath method combined with the over-relaxation method. Periodic boundary conditions were implemented for $N=L^2$  site clusters
with $L=12-60$ and $L=72,96$ for some calculations. The system is gradually cooled down from high temperature. A run at each magnetic field or temperature contains
typically $0.1-1\cdot10^6$  Monte Carlo steps (MCS's) for initial relaxation and twice MCS's during the calculation of mean values.

The different phases can be easily identified from the observation of real-space spin textures (e.g. see Fig. \ref{fig:SnapsN48}). We have also computed the average of various
physical quantities in order to identify precisely the  different phases and the corresponding transitions between them. We first calculate the standard  magnetization, magnetic susceptibility and specific heat.

\ba
M&=&\frac{1}{N}\left\langle\sum_{\vr}S^{z}_\vr\right\rangle,\quad \chi_z=\left\langle\frac{dM}{dh}\right\rangle,\\
\label{eq:MandChiOrderPara}
C&=&\frac{\langle E^2\rangle-\langle E\rangle^2}{NT^2}.
\label{eq:CvOrderPara}
\ea
We also introduce the discretized Skyrmion number  $\chi_Q$ and the total chirality\cite{Anderson, Nagaosa_NCL} $\chi_L$, defined as
\ba
\label{eq:ChiraOrderPara0}
\chi_Q&=&\frac{1}{4\pi}\left\langle\sum_{\vr_i}A^{(23)}_{\vr_i}\text{sign}[\chi^{(23)}_{L,\vr_i}]+A^{(45)}_{\vr_i}\text{sign}[\chi^{(45)}_{L,\vr_i}]\right\rangle\\
\label{eq:ChiraOrderPara}
\chi_L&=&\frac{1}{8\pi}\left\langle\sum_{\vr_i}\chi^{(23)}_{L,\vr_i}+\chi^{(45)}_{L,\vr_i}\right\rangle
\ea
where $A^{(ab)}_{\vr_i}=\|(\Sp_{\vr_a}-\Sp_{\vr_i})\times(\Sp_{\vr_b}-\Sp_{\vr_i})\|/2$ is the local area of the surface spanned by three
spins on every elementary triangle $\vr_i,\vr_a,\vr_b$ (see Fig. \ref{fig:lattice}b). Here $\chi^{(ab)}_{L,\vr_i}=\Sp_{\vr_i}.(\Sp_{\vr_a}\times\Sp_{\vr_b})$
is the so-called local chirality and $\vr_i,\vr_1\sim\vr_4$ are the sites  involved in the calculation
of $\chi_Q$ (see Fig. \ref{fig:lattice}). It is important to note that for slowly varying spin configurations $\chi_Q$ and $\chi_L$  would coincide.
In the case at hand the use of $\chi_Q$ turns out to be more effective because in
the discrete case it approximates much better the directed
area of the sphere surface, therefore providing a much better measure of the Skyrmion number.
We have computed the discretized Skyrmion number by sublattice and so the sites $\vr_i,\vr_a,\vr_b$ run in one sublattice.
However, the value of the local chirality without sublattice distinction is also an important quantity within the context of the anomalous Hall effect.
Indeed, electrons moving in the antiferromagnetic background of our system, and within the framework of the adiabatic approximation, would feel an
effective flux per plaquette which is given by $\chi_L$ \cite{Anderson, Nagaosa_NCL}, computed between sites of the three different sublattices.

\begin{figure}[htb]
\begin{centering}
\includegraphics[width=8.2cm]{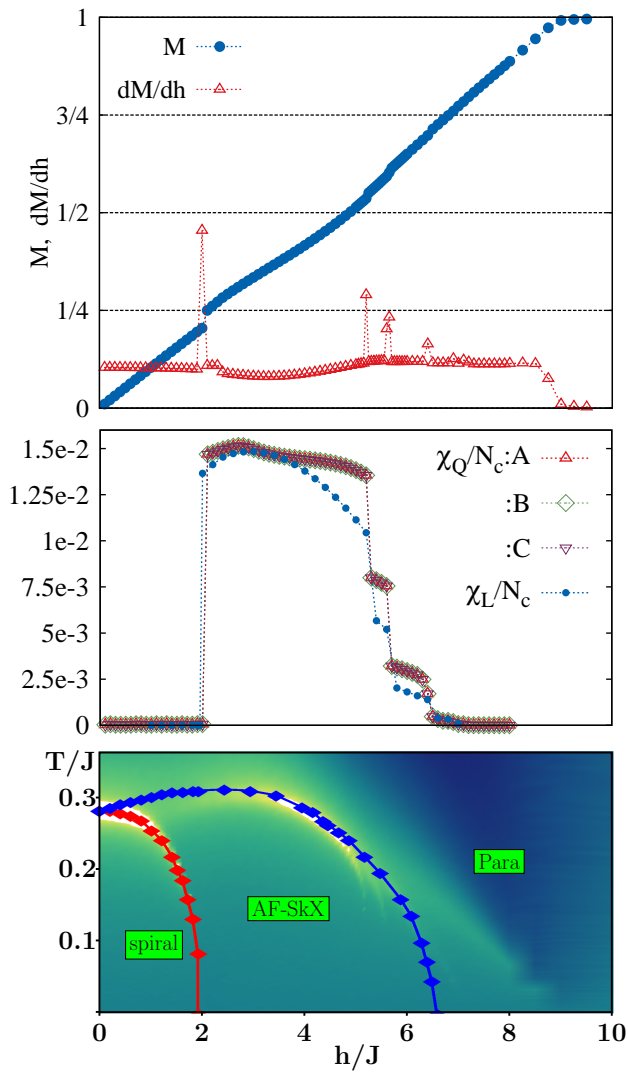}
\end{centering}
\caption{(Color online) Top: Magnetization $M$ (blue) and magnetic susceptibility vs. magnetic field $h$ for $L=48$, $D/J=0.5$ and $T/J=9\times10^{-3}$. Center:
Discretized Skyrmion number $\chi_Q$ and total chirality $\chi_L$, both quantities per sublattice, vs. $h/J$  for$L=48$, $D/J=1/2$. Red triangles, green rhombus and purple triangles
indicate the discretized Skyrmion number $A,B$ and $C$. Blue dots show the total chirality. Bottom:  Complete $h-T$ phase diagram. The limit of the  obtained by the peaks in
the specific heat and from the changes in the chirality order parameter.}
\label{fig:MvsH_Phasediagram}
\end{figure}

With the help of $\chi_Q$ we can detect the Skyrmion phase as a function of the magnetic field as shown in Fig. \ref{fig:MvsH_Phasediagram} (Center). At low magnetic field $h<h^*(T)$, $\chi_Q=0$
due to the spiral configuration. Then for $h^*(T)<h<h^{**}(T)$ a non zero valor of $\chi_Q$ indicates the presence of a wide AF-SkX phase. For instance,  in this region  and
for a system size of $N=48^2$  the discretized Skyrmion number in each sublattice $\chi_Q/N_c\sim 12/N_c=1.5\times10^{-2}$ ($N_c=N/3$) which is in perfect agreement with the number of Skyrmions that can be directly observed
from the MC snapshots (in Fig. \ref{fig:SnapsN48} only a portion of the total lattice is shown). We also plot $\chi_L$ together with $\chi_Q$ showing that $\chi_Q$ is a better measure of the Skyrmion number.  We observe that the discretized Skyrmion number is almost constant in the complete region and decay
in few steps as a consequence of the discrete nature of the Skyrmion patterns. In this region of the phase diagram all the spins have $z$-component greater than zero and its $xy$-components
look like an array of  2D-vortices, a pattern that can be interpreted as the remnant of the Skyrmion lattice. The stepwise decrease of the Skyrmion number
should be associated with the finite size of the lattice, which, due to the periodic boundary conditions imposed, can accommodate only definite numbers of
Skyrmions. In the continuum limit we expect that this number will smoothly go to zero.  In order to test this idea, we have performed an approximate
(continuum limit) computation of the Skyrmion radius as a function of the magnetic field, which leads to a smooth decrease of the Skyrmion density (see
Sec. \ref{sec-phenomelological}).

In all the MC simulations the spin configurations have a three sublattice structure, so in order to understand better the underlying structure, we plot the
patterns by sublattice, $A,B$ and $C$. As an example, in figure \ref{fig:SnapsN48} we show representative spin configurations for $L=48$ (only a small region
of the entire lattice is shown to illustrate the spin texture) at three different magnetic fields: $h/J=(i)1$, $(ii)2.4$ and $(iii)6.8$.

\begin{figure}[htb!]
\begin{centering}
\includegraphics[angle=90,width=8.5cm]{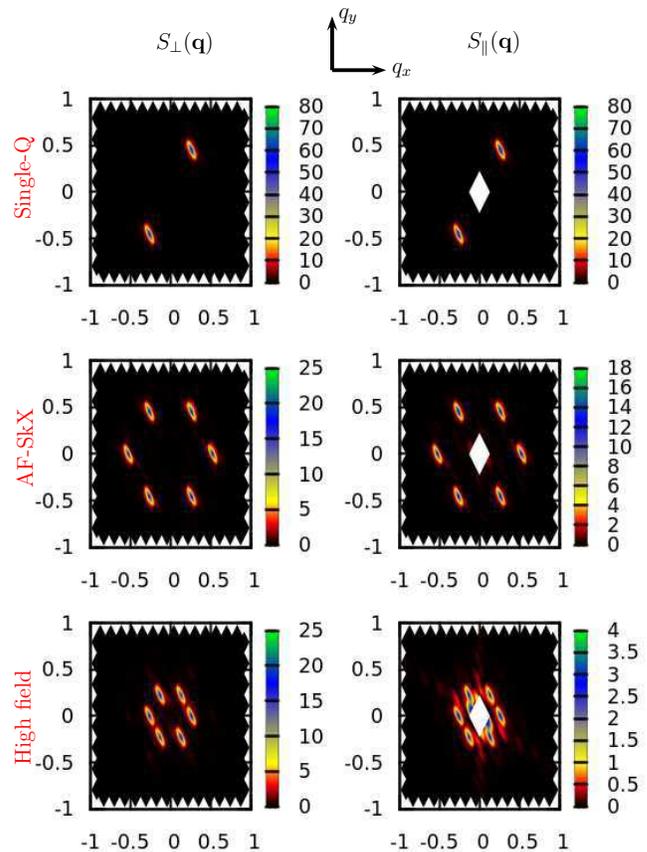}
\par\end{centering}
\caption{(Color online) Intensity plot of the static spin structure factor for $L=96$ for the spiral phase (Top); AF-SkX phase (Center) and disordered phase (Bottom).}
\label{fig:Strucfact}
\end{figure}

We have also calculated the static spin structure factor in the reciprocal lattice to identify the Bragg peaks that characterize the different spin-textures.
The perpendicular and parallel (to $z$) components $S_{\perp}(\vq)$ and $S_{\|}(\vq)$ are defined as
\ba
S_{\perp}(\vq)&=&\frac{1}{N}\langle|\sum_{\vr}S^{x}_{\vr}e^{-i\vq\cdot\vr}|^2+|\sum_{\vr}S^{y}_{\vr}e^{-i\vq\cdot\vr}|^2\rangle\\
S_{\|}(\vq)&=&\frac{1}{N}\langle|\sum_{\vr}S^{z}_{\vr}e^{-i\vq\cdot\vr}|^2\rangle
\ea
where $\langle\,\rangle$ means the averaged MC configurations. In figure \ref{fig:Strucfact} we show the intensity of the spin structure factor for the
patterns found (figure \ref{fig:SnapsN48}): spiral or single-q (Top) and AF-SkX or triple-q (Center) and high field phase (Bottom).

\begin{enumerate}
 \item[(i)] Spiral phase (Fig.\ref{fig:SnapsN48}a): the spin structure consists of three interpenetrating spirals on each sublattice, $A,B$ and $C$ (see figure \ref{fig:lattice}).
 Each one is characterized by one of the three possible ordering wave vectors $\vk^*$ obtained by the Spherical Approximation (Eq. \ref{eq:kc} and
 figure \ref{fig:wk-sphericalApp}). Both the $xy$ and $z$ components are characterized by the same wave vector as can be seen from the figure \ref{fig:Strucfact}.
%  In the ``wave front'' the spins are placed onto  a plane perpendicular to the direction of the propagation.

 \item[(ii)] Antiferromagnetic Skyrmion lattice phase (Fig.\ref{fig:SnapsN48}b): here the stable phase corresponds to three SkX phases, one on each sublattice. Each of the SkX
 phases is a superposition of three spirals characterized, both the $xy$ and $z$ components, by the three wave vectors $\vk^*$. The complete superposition
 of these three sublattice structures is shown in Fig.\ref{fig:SnapsN48}d.

\item[(iii)] High field phase (Fig.\ref{fig:SnapsN48}c) which in some areas shows a vortex-like structure (Fig.\ref{fig:SnapsN48}c). There, the
$xy$  components of the spins form vortices while the $z$ component is always positive.  This phase is established near the right edge (higher field sector) where the chirality order parameter, both $\chi_Q$, $\chi_L$  show smaller plateaux
\end{enumerate}


\begin{figure}[htb!]
\begin{centering}
\includegraphics[width=8.5cm]{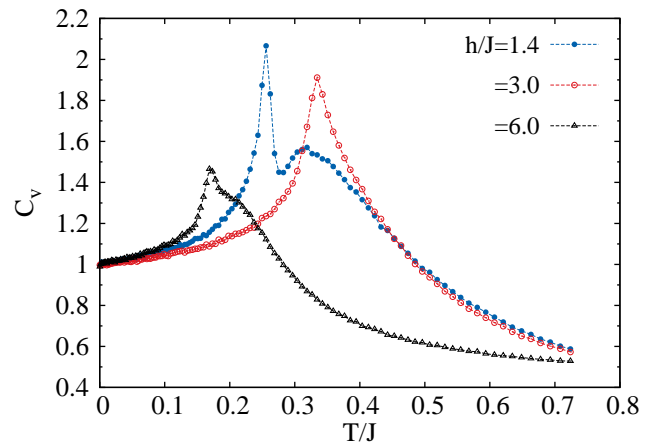}
\par\end{centering}
\caption{(Color online) Specific heat vs. temperature  for $h/J=1.4$, $3$ and $6$. The positions of the peaks are used to locate the lines separating the different phases.}
\label{fig:Cv}
\end{figure}

The main result of our study is summarized in the phase diagram shown in Fig. \ref{fig:MvsH_Phasediagram} (Bottom), which was obtained for $D/J=1/2$ (other
values of this ratio lead to a similar phase diagram). The AF-SkX phase is surrounded by the Spiral and paramagnetic phase and the lines separating the phases
were obtained in different ways, locating the peaks in the specific heat (figure \ref{fig:Cv}), the peaks in the magnetic susceptibility (Fig. \ref{fig:MvsH_Phasediagram}(top))
and using $\chi_Q$ as the order parameter (figure \ref{fig:MvsH_Phasediagram}(center)).

A very interesting behavior is observed when the system goes from the AF-SkX phase to a region where the $XY$ components of the spins form a vortex-like structure. In all the AF-SkX region, as we increase the magnetic field,
the Skyrmion number does not change.  To increase the magnetization, the Zeeman term  tends  to align  the spins, then in a big region of the  phase diagram,
antiferromagnetic exchange and  DM interactions favor the AF-SkX phase with a specific topological number (the number of Skyrmions). If we continue
increasing the magnetic field, the system (being discreet and in a phase with a given number of Skyrmions) cannot increase the magnetization without paying a high
energetic cost, thus what is observed is that the AF-SkX phase changes the number of Skyrmions. This result is shown in figure \ref{fig:MvsH_Phasediagram}(Center)
where both the discretized Skyrmion density and total chirality density decrease by steps.

In the next Section we use a phenomenological-variational analysis to compute the dependence of the Skyrmion radius with the magnetic field.

\section{Phenomenological analysis of the Skyrmion lattice spacing}
\label{sec-phenomelological}
Although in the MC data the number of Skyrmions for a given system size, and hence the lattice spacing varies with the applied magnetic field by
discrete jumps, we suspect this behavior to be a finite size effect. In the thermodynamic limit one can expect a lattice spacing continuously varying
as a product of the competition between the applied magnetic field which tends to increase the magnetization by making bigger Skyrmions and the DM
interaction for which a fixed Skyrmion size is optimal. To model this competition, we follow the same idea developed by Jung Hoon Han {\it et. al} \cite{Nagaosa2010}
in the case of a ferromagnetic system. Of course here we have a three sublattice Skyrmion pattern so the phenomenological description of a free energy
in terms of spin orientations that we present has to be interpreted as representative of an average free energy for one of the three sublattices.

Consider the local spin orientation $(\theta,\phi)$ of a single Skyrmion depending on the local coordinate $(r,\phi)$ as $\phi=\varphi-\pi/2$ and $\theta=\theta(r)$. The total free energy of a single Skyrmion then reads
\vspace{1cm}

\ba
F_{Sk}&=&4\pi J\int\,r dr\left[\left(\frac{1}{2}\frac{d\theta}{dr}+\kappa\right)^2-\kappa^2+\frac{\kappa}{2r}\sin(2\theta)\right.\nn\\
&&\left.+\frac{\sin^2(\theta)}{4r^2}-\eta(\cos(\theta)-1)\right]
\label{eq:Fsk}
\ea
where $\eta=h/(2J)$ and $\kappa=D/(2J)$.
The function $\theta(\vr)$ minimizing the functional (\ref{eq:Fsk}) describes the magnetization distribution in an isolated Skyrmion. To understand
the observed transitions phenomenologically, we calculate the energy of a isolated Skyrmion using the following trial function
\ba
\theta(r)&=&\pi, \quad r\leq R_0\\
&=&\pi\frac{1-\frac{r-R_0}{R-R_0}}{1+b\frac{r-R_0}{R-R_0}}, \quad R_0< r\leq R\\
&=&0, \quad R<r
\label{eq:thetar}
\ea

\begin{figure}[htb!]
\includegraphics[width=8cm]{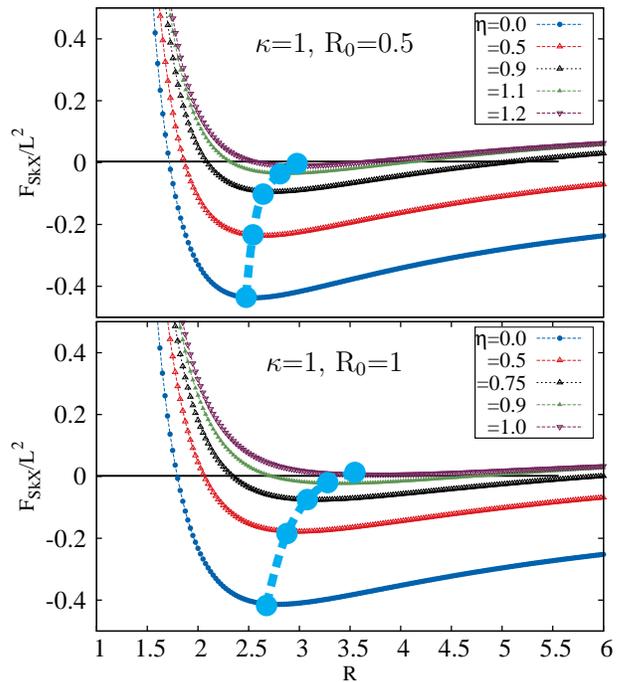}
\caption{(Color online) Dependence of the free energy density given by Eq. (\ref{eq:FSkX}) as a function of the Skyrmion radius $R$ for two values of the Skyrmion core: $R_0=0.5,1$ ($\kappa=1$) }
\label{fig:FvsR}
\end{figure}

In the definition of $\theta(r)$, the parameter $R_0>0$ is the minimum radius simulating the central spin in the continuous limit, $R>R_0$ is the
radius of the Skyrmions and $b$ is a parameter to be computed self-consistently. The parameters $R,b$   are fixed by the condition of minimum, $\{dF_{SkX}/dR,dF_{SkX}/db\}=\{0,0\}$.
The Skyrmion lattice observed in our simulations (in each sublattice) correspond to a close-packing of individual Skyrmions
in a triangular lattice. We then write the total free energy of the Skyrmion lattice state as

\ba
F_{SkX}&=&3\frac{L^2}{2\sqrt{3}R^2}F_{Sk}
\label{eq:FSkX}
\ea
where $L$ is the size system of the system.
In Fig. \ref{fig:FvsR} we show a few plots of  $F_{SkX}$ as a function of parameter $R$, i.e. as a function of the Skyrmion radius. We observe that
the behavior of the optimal Skyrmion spacing as a function of the magnetic field varies very slowly for small values of the magnetic field and
then behaves much more abruptly when approaching the high field paramagnetic phase. In a fixed finite size sample, where the Skyrmion lattice has
to be commensurate with the sample size, this translates precisely in a large  zone of the phase diagram in which the Skyrmion number is fixed
and then start to produce jumps that appear more and more abruptly.

%

 %%%%%%%%%%%%%%%%%%%%%%%%%%%%%%%%%%%%%%%%%%%%%%%%%
 \section{Skyrmion persistence and the importance of the lattice}
 \label{sec-OutEquil}
 %%%%%%%%%%%%%%%%%%%%%%%%%%%%%%%%%%%%%%%%%%%%%%%%%

\begin{figure}[htb]
   \centering{}
 \includegraphics[width=8.9cm]{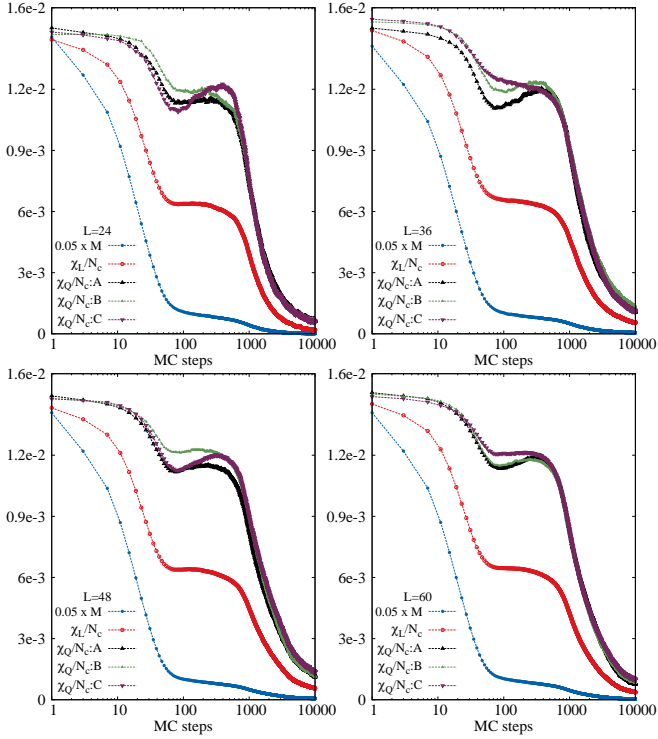}
 \caption{(Color online) The Monte Carlo time evolution of the magnetization, discretized Skyrmion number $\chi_Q$ and total chirality $\chi_L$ defined by Eqs. (\ref{eq:ChiraOrderPara0}) (\ref{eq:ChiraOrderPara}),
 calculated at $T = 9\times10^{-3}\,J$ for sizes of $L =24,36,48,60$. The initial state corresponds to the AF-SkX phase (Fig. \ref{fig:SnapsN48}). Each curve corresponds to an average of 500 independent time evolutions.}
 \label{fig:ChiralityvsMCsteps}
 \end{figure}

It is  well known that Skyrmions are topological defects with an infinite life-time in the continuum and with a very long persistence time in ferromagnetic lattice models.
Here we are going to argue that despite having an apparently more fragile three sublattice structure, the Skyrmion persistence is also quite noticeable.
To do this we start from a configuration, like figure \ref{fig:SnapsN48} in the AF-SkX phase,
obtained by annealing at fixed magnetic field. In Fig. \ref{fig:ChiralityvsMCsteps}, we show an example of the MC time evolution (the curves are the average of $500$ copies) of such initial state in our simulations after a quench to zero magnetic field. Note that unlike what is observed with the normalized magnetization $M$ ($\times 0.05$ to use the same scale for all quantities), the total chirality $\chi_L$ and discretized Skyrmion number by sublattices $\chi_Q$ exhibit greater stiffness extending over several decades of time.

To complete the analysis we have also performed MC simulations in the non-frustrated square lattice  where in contrast to the case of the triangular
lattice, $\delta \hat{\vr}=(\vr'-\vr)/|\vr'-\vr|$ is an unitary vector pointing along the axis  $\vec{e}_1=(1,0)$ and $\vec{e}_2=(0,1)$. Our
simulations do not show any evidence of an antiferromagnetic Skyrmion lattice in the square lattice case. One possible explanation for the absence of this phase is geometrical frustration. At the microscopic level,
on the square lattice, spins have the tendency to form a two sublattice structure, instead of three for the triangular lattice. On the other
hand, at the Skyrmion lattice scale, the energetically favored lattice is triangular, which fits perfectly well with a three sublattice structure
but produces frustration in the two sublattice case (if one sublattice arranges in a triangular array, the other sublattice has to fit in an hexagonal array, which is energetically less
favorable). As such, and in contrast to the ferromagnetic interactions case, the choice of the lattice seems crucial  for the formation of the SkX phase.

%%%%%%%%%%%%%%%%%%%%%%%%%%%%%%%%%%%%%%%%%%%%%%%%%
\section{Conclusions}
\label{sec-Conclusions}
%%%%%%%%%%%%%%%%%%%%%%%%%%%%%%%%%%%%%%%%%%%%%%%%%
In this paper we have studied the triangular lattice antiferromagnet with classical spins in the presence of inversion symmetry breaking interactions
and in the presence of an external magnetic field, in order to investigate the possible appearance of Skyrmion lattice structures in a pure antiferromagnetic system.
We have run extensive Monte Carlo simulations complemented by simple analytical techniques.
The most interesting outcome is the stabilization of a novel state, which we term "antiferromagnetic Skyrmion lattice" (AF-SkX), which consists of three
interpenetrating SkX states of the usual type observed in many ferromagnetic models. This phase is stable for a wide range of magnetic fields and couplings.

Observing the evolution of the topological order parameter as the magnetic field increases, which we define as the chirality per sublattice and which can be in turn
associated to the topological charge or number of Skyrmions, we observe a clear jump, indicating a discontinuous transition from a
spiral phase into the AF-SkX phase, where multiple Bragg peaks coexist in the spin structure factor. For higher fields, a second (probably continuous)
transition occurs into a featureless paramagnetic phase.

We also observe that after a magnetic field quench, the Skyrmion number persists for longer MC times as one should have expected, making apparent the
robustness of the AF-SkX state. The relation of this AF-SkX state with the 3-sublattice structure of the lattice studied in the present paper is
contrasted with the square lattice case, where such a state does not show up. Finally, we have also shown that doped electrons that would evolve in
this background, would feel a fictitious non-zero magnetic flux, as depicted in the central figure of \ref{fig:MvsH_Phasediagram}. Interestingly,
the total chirality, which is directly related to the local flux felt by the doped electrons, also remains quite constant in agreement with the fact that the AF-SkX structure
is remarkably stable for a wide range of magnetic fields, and this of course can have very interesting consequences in the context of the anomalous Hall effect.

%%%%%%%%%%%%%%%%%%%%%%%%%%%%%%%%%%%%%%%%%%%%%%%%%
\section*{Acknowledgments}
%%%%%%%%%%%%%%%%%%%%%%%%%%%%%%%%%%%%%%%%%%%%%%%%%
PP acknowledges Tsuyoshi Okubo for enlightening discussions. HDR and DCC are partially supported by
PIP 0747 CONICET and PICT 1724.

\bigskip

\appendix
\section{Matrix Elements for Spherical approximation}
\label{AppendixA}

The explicit expressions for the $3\times3$ matrices of the matrix (\ref{eq:matrixK}) are given by
\ba
m_{12}(\vk)=[J_{12}^{\alpha\beta}],m_{13}(\vk)=[J_{13}^{\alpha\beta}],m_{32}(\vk)=[J_{32}^{\alpha\beta}]
\ea

where

\ba
J_{12}^{\alpha\beta}&=&(1+\gamma^{*}_{1,\vk}+\gamma^{*}_{2,\vk})\delta^{\alpha\beta}\nn\\
&&+\kappa\sum^{2}_{\gamma=1}(\vec{e}_1-\gamma^{*}_{1,\vk}\vec{e}_2-\gamma^{*}_{2,\vk}\vec{e}_3)^{\gamma}\varepsilon^{\alpha\beta\gamma}\\
J_{13}^{\alpha\beta}&=&(1+\gamma^{*}_{1,\vk}\gamma_{2,\vk}+\gamma^{*}_{1,\vk})\delta^{\alpha\beta}\nn\\
&&+\kappa\sum^{2}_{\gamma=1}(\vec{e}_2+\gamma^{*}_{1,\vk}\gamma_{2,\vk}\vec{e}_3-\gamma^{*}_{1,\vk}\vec{e}_1)^{\gamma}\varepsilon^{\alpha\beta\gamma}\\
J_{32}^{\alpha\beta}&=&(1+\gamma_{1,\vk}\gamma^{*}_{2,\vk}+\gamma^{*}_{2,\vk})\delta^{\alpha\beta}\nn\\
&&+\kappa\sum^{2}_{\gamma=1}(\vec{e}_3+\gamma_{1,\vk}\gamma^{*}_{2,\vk}\vec{e}_2-\gamma^{*}_{2,\vk}\vec{e}_1)^{\gamma}\varepsilon^{\alpha\beta\gamma}
\ea

where $\vec{e}_1=(1,0)$, $\vec{e}_2=(1/2, \sqrt(3)/2)$, $\vec{e}_3=\vec{e}_1-\vec{e}_2$ and $\gamma_{1,\vk}=e^{-i\vk\cdot(\vec{e}_1+\vec{e}_2)}$ and $\gamma_{2,\vk}=e^{-i\vk\cdot(2\vec{e}_1-\vec{e}_2)}$.

\end{document}